\documentclass[twocolumn,showpacs,pre,amsmath,amssymb,superscriptaddress,nofootinbib]{revtex4}
\usepackage{amsmath,amssymb,bm,graphicx,epsfig,psfrag,ulem,color,natbib}
\usepackage[abc]{overpic}
\newcommand{\bx}{\mathbf {x}}
\newcommand{\by}{\mathbf {y}}
\newcommand{\bk}{\mathbf {k}}
\newcommand{\bv}{\mathbf {v}}
\newcommand{\bvf}{\mathbf{v}_{f}}
\newcommand{\etal}{{\it et al.~}}
\begin{document}

\title{\bf Model flocks in a steady vortical flow}

\author{A.~W.~Baggaley}
\affiliation{School of Mathematics and Statistics, Newcastle
University, Newcastle upon Tyne, NE1 7RU, UK}\affiliation{Joint Quantum Centre Durham-Newcastle}


\begin{abstract}
\noindent
We modify the standard Vicsek model to clearly distinguish between intrinsic noise due to imperfect alignment between organisms, and extrinsic noise due to fluid motion. 
We then consider the effect of a steady vortical flow, the Taylor Green vortex, on the dynamics of the flock, for various flow speeds, with a fixed intrinsic particle speed. We pay particular attention to the morphology of the flow, and quantify its filamentarity.  Strikingly, above a critical flow speed there is a pronounced increase in the filamentarity of the flock, when compared to the zero flow case. This is due to the fact that particles appear confined to areas of low vorticity; a familiar phenomena, commonly seen in the clustering of inertial particles in vortical flows. Hence the  cooperative motion of the particles gives them an effective inertia, which is seen to have a profound effect on the morphology of the flock, in the presence of external fluid motion. Finally we investigate the angle between the flow and the particles direction of movement and find it follows a power law distribution.
\end{abstract}

\pacs{47.32.Ef,47.63.-b,87.18.-h}
\maketitle

\section{Intoduction}\label{sec:intro}
\noindent
Flocks of birds, schools of fish or swarms of insects are all examples of self-organised aggregations, offering some of the most spectacular displays of motion in nature. The complexity of the animals moving together, seemingly as one, has intrigued observers for centuries. Indeed, the ability of a  flock of birds to change direction apparently simultaneously and spontaneously led Edmund Selous, who published numerous books on natural history, to believe birds to be telepathic \cite{selous1931thought}.

To mathematically model animal aggregations it is typical to use models based on self-propelled particles, such as the Vicsek model \cite{Vicsek_1995}. Here the direction of motion is typically based on the idea that flocking animals tend to align themselves to others within a local ``radius of interaction'' or based on a number of nearest-neighbours, as observed in starling flocks \cite{Young_2013}. More complex models with additional short range repulsion have also been investigated \cite{Couzin2002}, and a recent study extended the Vicsek model to include
decision making, needed for example by birds to make a synchronised landing \cite{Bhattacharya2010}.

We note that this is not the only approach with a number of interesting studies based on nonlinear partial differential equations \cite{Parrish1999}. In addition a recent study by Ballerini \etal \cite{Ballerini2008} questioned the validity  of approaches based on the standard Vicsek model, arguing that the radius of interaction should be based on a topological distance, rather than the metric distance used in \cite{Vicsek_1995}. However we believe that the standard  Vicsek model \cite{Vicsek_1995} is the ideal starting point to investigate the effect of external fluid motions on flock dynamics and morphology, and note that the behaviour of marching locusts could be modelled using a similar approach \cite{Buhl2006}. There are also interesting extensions to model human crowd stampedes \cite{Helbing_2000}.

One of the drawbacks of the standard Vicsek model is that all sources of noise in the system arise in a single term, which is both temporally and spatially uncorrelated. It is perhaps more natural to distinguish between two sources of noise, which we would expect to arise in such systems. The first is intrinsic noise, due to the fact that animals will never perfectly align. Here we see no issue with approach used in \cite{Vicsek_1995} where uncorrelated noise seems an appropriate approximation. However animals move in fluid environments, and so will be subject to external noise, due to the motion of the fluid. In almost all natural flows a combination of large lengthscales and small viscosities mean that the flow is turbulent \cite{uriel1995turbulence}. Hence the external forcing will be subject to complex spatio-temporal correlations which are not present in the standard Vicsek model.

A recent study by Khurana \& Ouellette \cite{Khurana2013}  sought to address this issue by considering a slight modification to the Vicsek model, in an unsteady flow made up of a summation of random Fourier modes, with a prescribed Energy spectrum consistent with the Kolmogorov spectrum of classical turbulence. The focus of \cite{Khurana2013} was the stability of flocks, particularly in the limit of a small radius of interaction, where they showed an external forcing with spatio-temporal correlations was far more efficient in destabilising the flock. Here our interest is in the effect of fluid motion on coherent flocks, with larger values for the radius of interaction.

\begin{figure*}[!ht]
\begin{center}
\includegraphics[width = 0.4\textwidth]{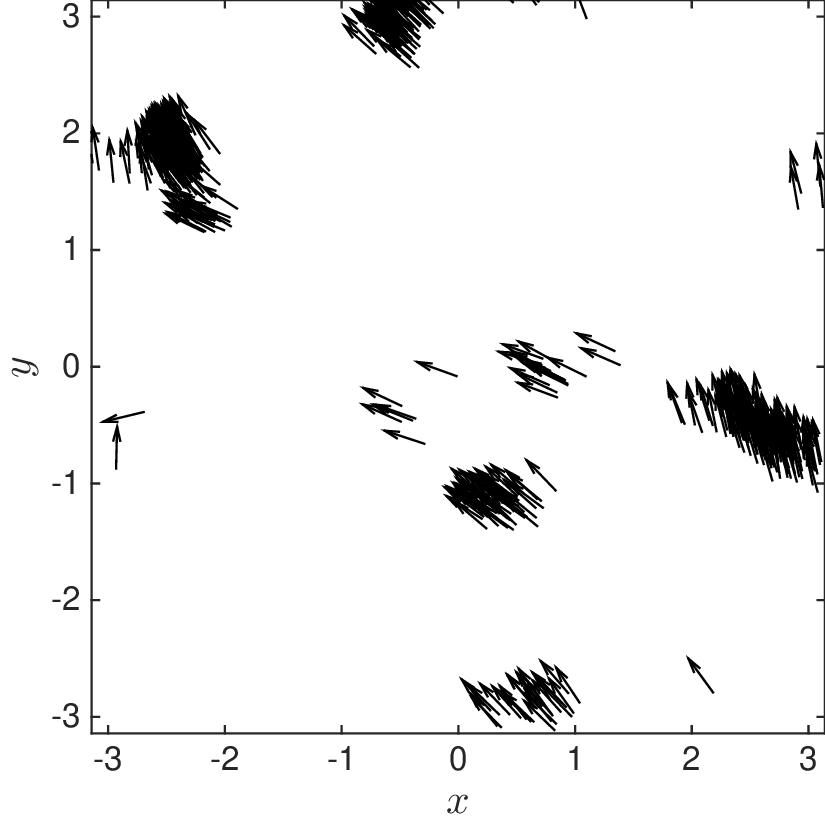}\hspace{1.75cm}
\includegraphics[width = 0.4\textwidth]{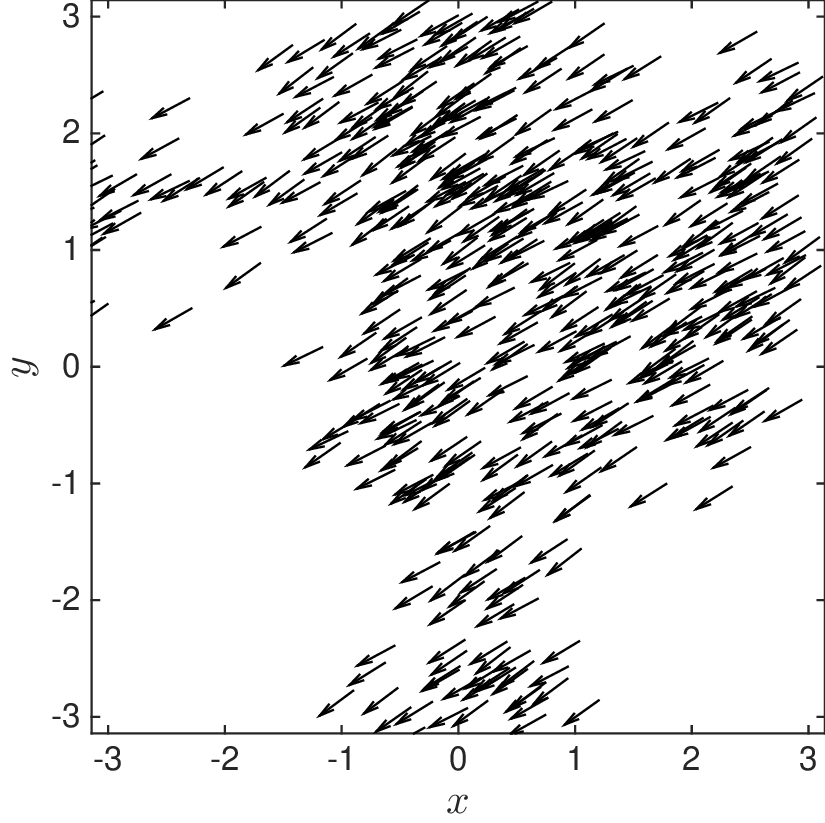}
\caption{Snapshots of the Vicsek model in the statistically steady-state (left $\eta=0.1, \, R=0.2$, right $\eta=0.1,\, R=0.8$). Arrows indicate the particles direction of motion.}
\label{fig:vicseck1}
\end{center}
\end{figure*}

One may be tempted, therefore, to surmise that it is then necessary to consider model flocks in fully turbulent flows. However,  implementing a full direct numerical simulation of the Navier-Stokes equation, necessary to produce a turbulent velocity field is a formidable computational challenge \cite{Ishihara2009}. Moreover interpretation of results is not necessarily straightforward or unambiguous, given the strongly nonlinear nature of fully developed turbulence. Even the `reduced' KS model considered in \cite{Khurana2013} is beyond what we wish to consider here, and it fails to reproduce many of the important features of true turbulence, such as intense vortical structures \cite{She1990}. There is a long tradition in fluid dynamics of using simple `toy' models, to probe the underlying physics of  a  problem, with a certain level of control. This is the approach we take here, turning to the Taylor-Green vortex \cite{Taylor1923} as an ideal toy flow.  Indeed this flow has been applied to fields as diverse as swimming microorganisms \cite{Durham2011}, the settling of inertial particles \cite{Maxey1986, Bergougnoux2014} and  Magnetohydrodynamics \cite{Nore1997}.

The aim of this paper is to investigate the effect of a simple vortical flow on a Vicsek flock, paying particular attention to the flock morphology. We note that the shapes of schools of fish and flocks of birds have received both theoretical \cite{Hemelrijk2010,Hemelrijk2012} and experimental attention \cite{Partridge1980,Hemelrijk2010}. However less attention has focused on the role that external fluid motion could play in affecting flock morphology. 

The plan of the paper is as follows: In Section \ref{sec:vicsek} we describe the standard Vicsek model. We also introduce statistical measures of the shape and structure of the flock, namely Ripley's K function and Minkowski functionals. In Section \ref{sec:TGV} we extend the model to include a smooth external forcing due to a steady vortical flow. Detailed comparisons of the statistical measures of shape and structure are made and we also study the angle between the particles and the flow. We close with a discussion of the results in Section \ref{sec:discuss} and highlight possible directions for future studies.

\begin{figure}
\begin{center}
\includegraphics[width = 0.4\textwidth]{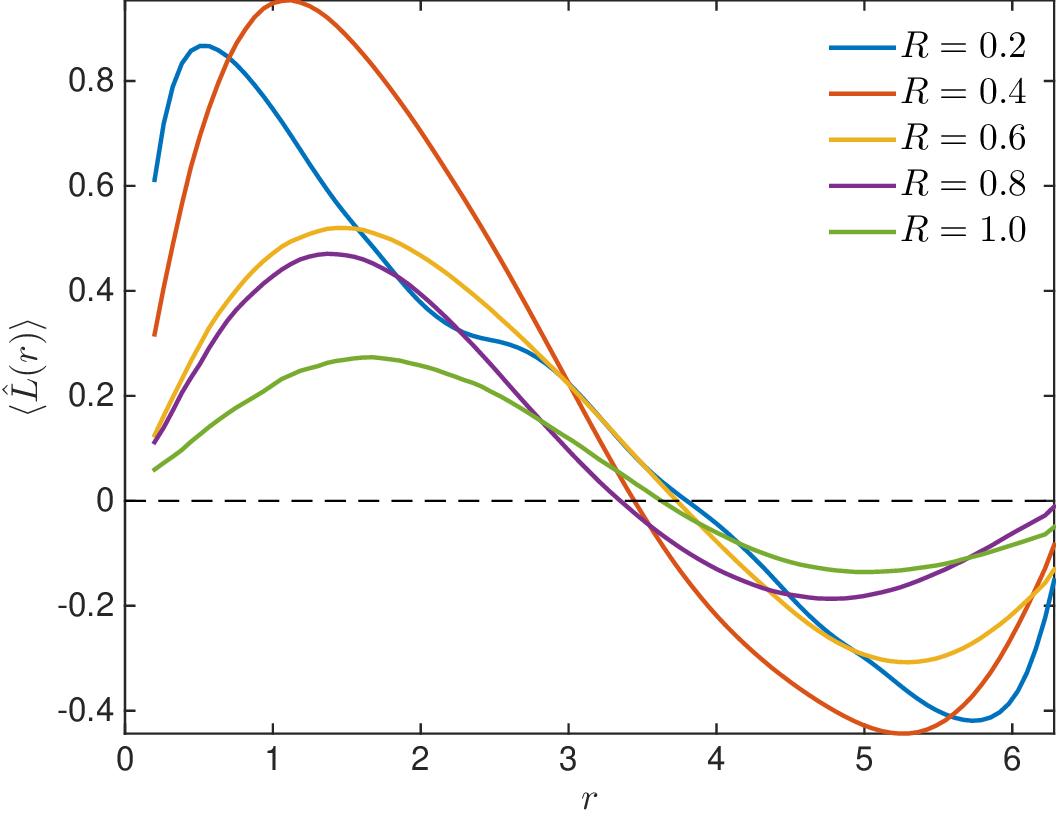}
\caption{(Color online) The time averaged Besag's function $\langle \hat{L}(r) \rangle $, Eq.~(\ref{eq:besag}) plotted as a function of spatial scale $r$ for simulations with $\eta=0.2$. $ \hat{L}(r) >0$ indicates clustering of particles at a given scale, $ \hat{L}(r) <0$ indicates segregation.}
\label{fig:ripley1}
\end{center}
\end{figure}

\section{The Vicsek model }\label{sec:vicsek}
\noindent
We begin by introducing the Vicsek model \cite{Vicsek_1995}, a well-known analogue of the Ising model of ferromagnetism. We consider $N$ particles in a two-dimensional square box (with sides of size $L$), with periodic boundary conditions. Whilst the systems we are interested in are three-dimensional, by working in a low dimensional space we gain a huge computational speedup which allows for a thorough examination of parameter space. A follow up study to investigate flocking in three-dimensional time dependent flows is planned in the near future. Each particle has a position $\bx_i(t)$ and an intrinsic, self-driven, velocity $\bv_i(t)$. All particles are assumed to move with the same speed, $V$, and a particles intrinsic velocity is determined by
\begin{equation}\label{Eq:velocity}
\bv_i=(V\cos(\theta_i),V\sin(\theta_i)),
\end{equation}
where $\theta_i$ determines the direction the particle moves in. Key to the model is that $\theta_i$ is periodically (at each time increment) determined from the average of the particle's own direction, plus the directions of its neighbours within a critical radius, $R$, such that \cite{atan2}
\begin{equation}\label{Eq:theta}
\theta_i={\rm atan2} \left( \frac{1}{N}\sum_{|\bx_i-\bx_j|< R} \sin \theta_j,  \frac{1}{N}\sum_{|\bx_i-\bx_j|  < R} \cos \theta_j \right) +\eta \zeta_i.
\end{equation}
The final term in  Eq.~(\ref{Eq:theta}) is used to model both intrinsic and extrinsic noise in the system. Specifically $\zeta_i$ is a uniformly distributed random variable on the interval $[-1,1]$ and $\eta$ is the intensity of the noise.
From these simple rules Vicsek \etal \cite{Vicsek_1995} observed the emergence of collective flocking or swarming of the particles. The global order of the system can be characterised by computing,
\begin{equation}
\psi_{\eta, R}(t)=\dfrac{1}{NV}\left | \sum_{i=1}^N \bv_i \right |,
\end{equation}
which is a convenient measure to establish if the system has reached a statistically steady state.
We follow \cite{Vicsek_1995} and model a system of $N=500$ self-propelled particles in a domain of size $L=2\pi$ with a speed $V=1$. Particles are evolved according to an Euler scheme such that
$$
\bx_i(t+\Delta t)=\bx_i(t)+\Delta t \bv_i(t)
$$
where at each tilmestep $\bv_i(t)$ is updated according to Eqns.~(\ref{Eq:velocity}) \& (\ref{Eq:theta}). Note $\zeta_i$ is drawn randomly at each timestep and we take $\Delta t=0.1$. For large value of $\eta$ cooperative motion is not observed, as the system is dominated by noise, hence we restrict our parameter values to $\eta \in [0,0.5]$, $R \in [0,1]$. In order to be as computationally efficient as possible all distance evaluations are computed using a $k$-d tree \cite{Bentley_1975}.

Simulations are evolved for 200 time units. Within a short period of time ($\approx 20$ time units) $\psi_{\eta, R}$ saturates to a statistically steady value, which for the low noise intensity considered here is close to or equal to unity, indicating substantial global alignment of the particles. Snapshots of the particles for simulations with $\eta=0.1$, $R=0.2, \, 0.8$ are displayed in Fig.~\ref{fig:vicseck1}, the effect that $R$ plays in determining the structure of the flock is clear.

In order to study this effect more quantitatively we use Ripley's K function, a statistical pattern analysis method used as a measure of spatial clustering. For $N$ particles this is defined as
\begin{equation}
\hat{K}(r) = \frac{L^2}{N}\sum_{i\ne j} \dfrac{I(|\bx_i-\bx_j|< r)}{N-1},
\end{equation}
where $I$ is the indicator function ($1$ if the Euclidean separation is less than $r$, $0$ otherwise), and $r \in [0,L]$. Note that this can easily be adapted to account for the periodic boundary conditions used in this study. Particle clustering results in $\hat{K}(r)$ increasing faster than if particles are distributed in a spatially random manner, that is, if they follow a Poisson distribution. Ripley's K function for a Poisson-distributed data set takes the form $\hat{K}(r) = \pi r^2$. For a linear scaling of Poisson-distributed data, it is common to normalise $\hat{K}(r)$ and define Besag's function as
\begin{equation}\label{eq:besag}
\hat{L}(r) =\sqrt{K(r)/\pi}-r.
\end{equation}
The advantage of Besag's function is that it has a simple interpretation, it is zero for randomly distributed points, takes positive values for particles clustered over the spacial scale $r$ and is negative if the particles are dispersed over a given scale. To increase our statistics we apply temporal averaging, and use angled brackets $\langle \rangle$ to denote temporal averaging in the statistically steady state (where $\psi_{\eta, R}$ has saturated); this is done every 10 time-steps. Figure \ref{fig:ripley1} shows the averaged  Besag's functions for five different values of interaction radius $R$ ($\eta=0.2$). A noticeable migration of the peak of $\langle \hat{L} \rangle$ to the right is observed, as well as a broadening of the values of $r$ where $\langle \hat{L} \rangle >0$. Both are consistent with clustering of particles over larger scales with increasing $R$. We want to emphasise that this is not a particularly novel result, however it is important to establish `baseline' statistics of the model, before studying the effects of flocking in the presence of a structured external flow.

\begin{figure}
\begin{center}
\includegraphics[width = 0.3\textwidth]{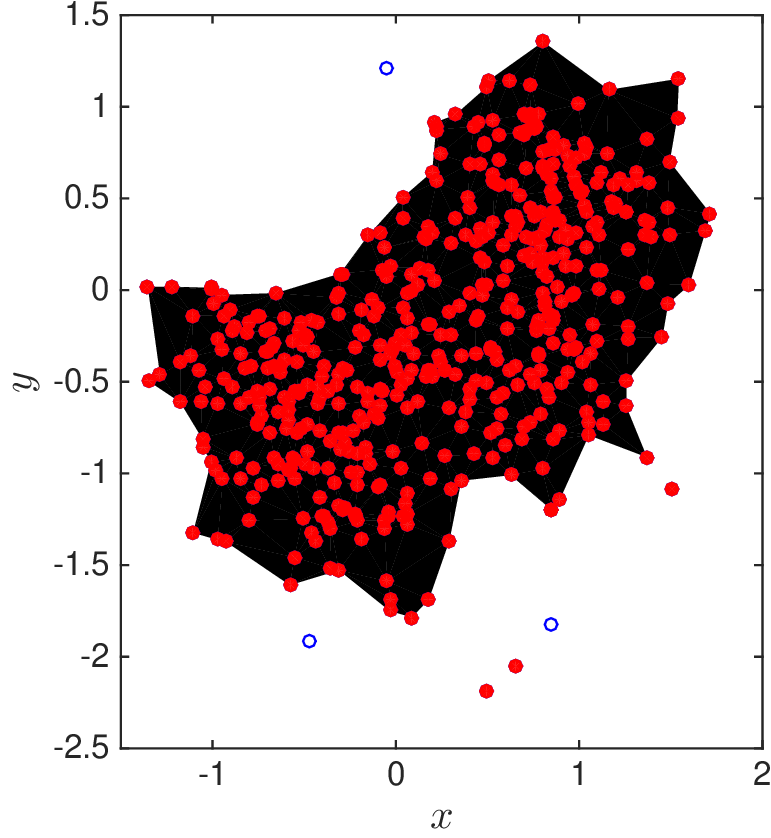}
\caption{(Color online) An example of a flock formed from the standard Vicsek model. Red (filled) points indicate where the separation between that point and its nearest neighbour is less than $\ell/2$, where  $\ell=L/\sqrt{N}$ is the expected separation between random points. Blue (open) points are those which do not satisfy this criteria. The black structure shows the $\alpha$-shape \cite{Edelsbrunner_1983} constructed from the red points, which is used to quantify the morphology of the flock.}
\label{fig:alpha}
\end{center}
\end{figure}

\begin{figure}
\begin{center}
\includegraphics[width = 0.4\textwidth]{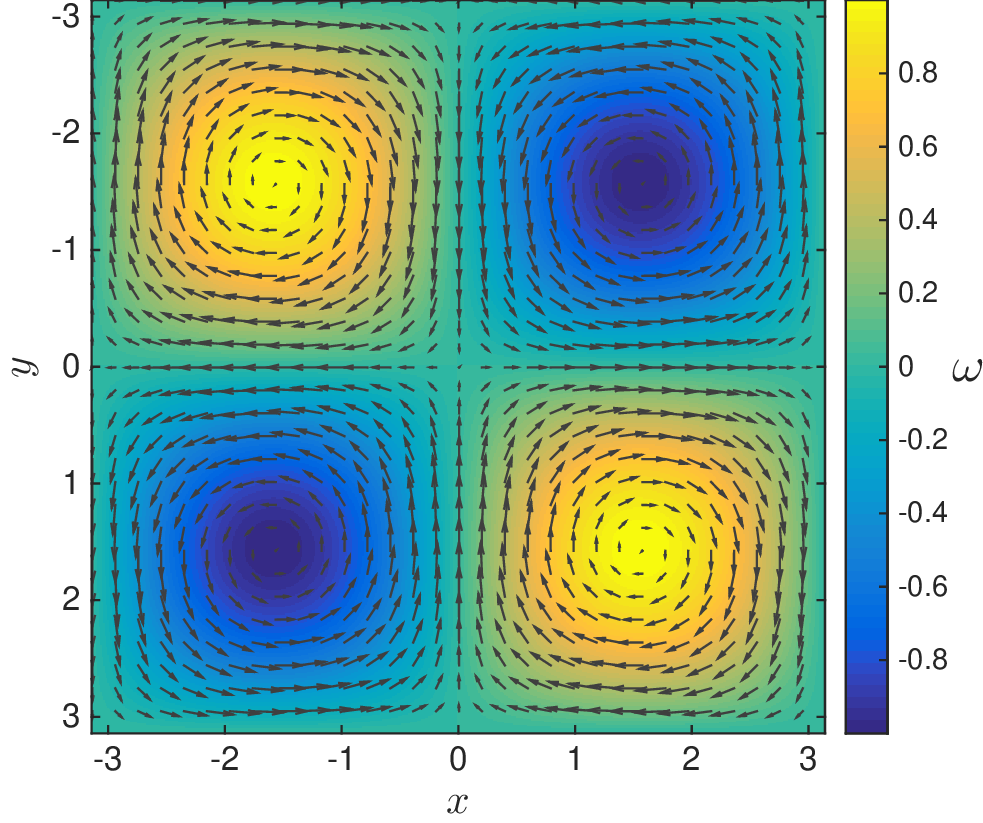}
\caption{(Color online) The Taylor-Green vortex flow. The pseudocolour plot displays the flow's vorticity, Eq.~(\ref{eq:TGVw}), black arrows indicate the velocity field, Eq.~(\ref{eq:TGV}).}
\label{fig:flow}
\end{center}
\end{figure}

\begin{figure*}[!ht]
\begin{center}
\includegraphics[width = 0.4\textwidth]{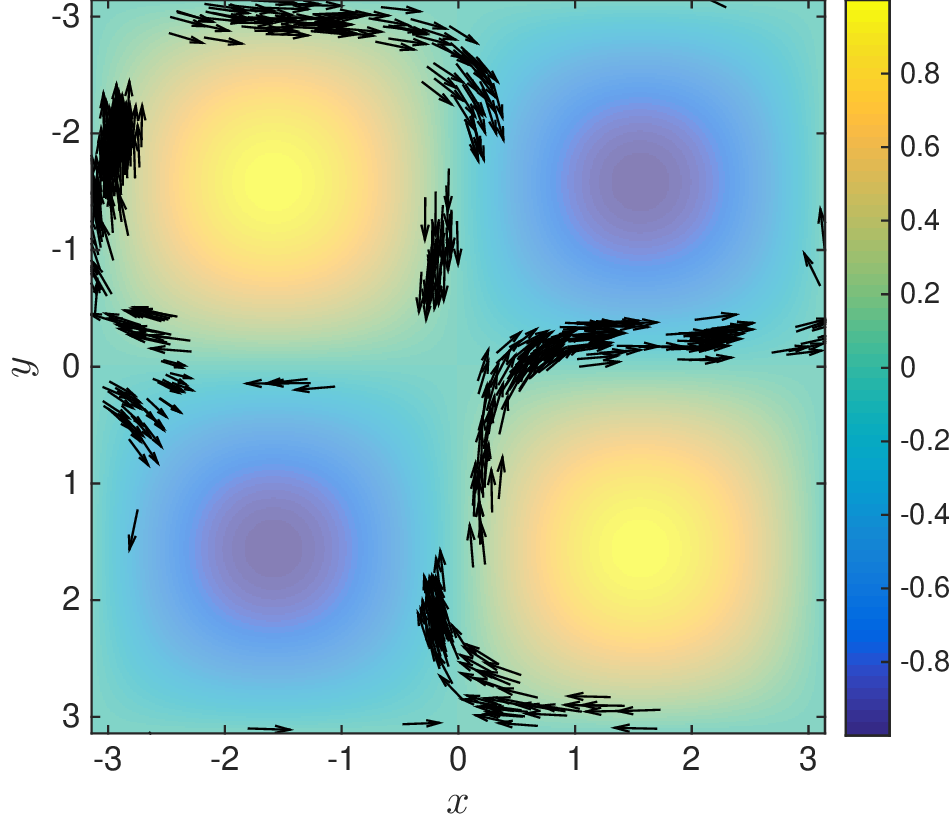}\hspace{1.75cm}
\includegraphics[width = 0.4\textwidth]{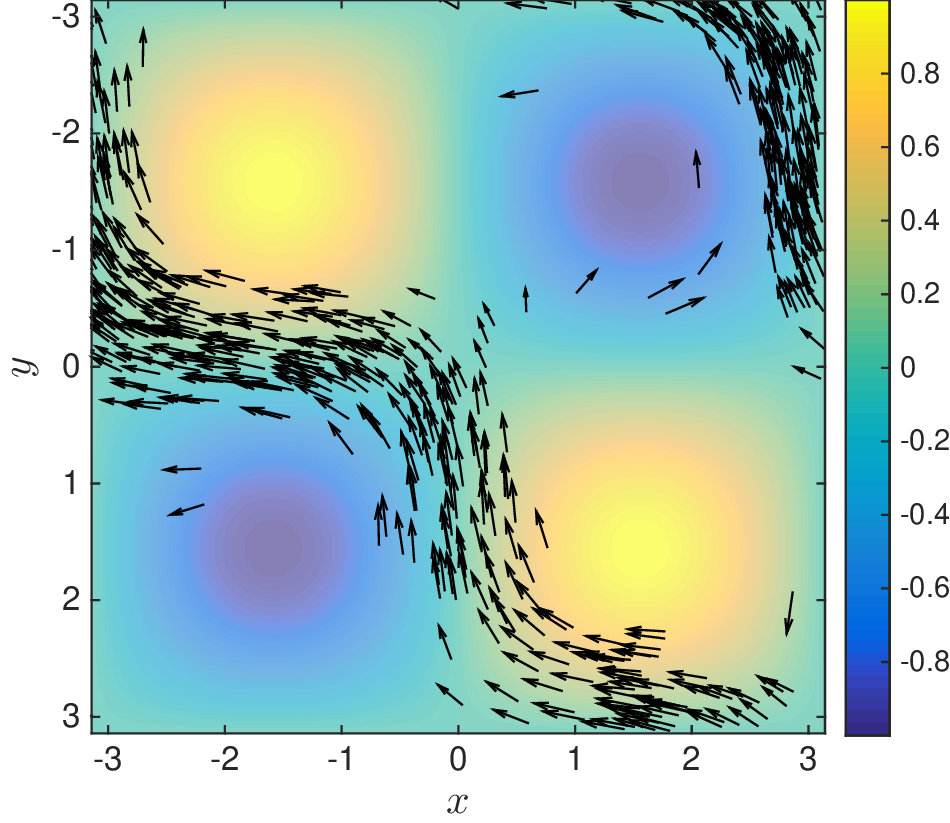}
\caption{(Color online) Snapshots of the Vicsek flock with external fluid forcing, in the statistically steady-state (left $\eta=0.1, \, R=0.2$, right $\eta=0.1,\, R=0.8$) with $V_f=0.5$. A pseudocolour plot (made slightly translucent to improve visualisation of the particles) displays the flow's vorticity $\omega$, Eq.~(\ref{eq:TGVw}). Arrows indicate the particles direction of motion. Note the concentration of particles in regions where $\omega \simeq 0$.}
\label{fig:vicseck2}
\end{center}
\end{figure*}

It is also instructive to introduce a statistic which measures the characteristic size of the flock. Whilst there is no unique way to do this, we find it convenient to measure
\begin{equation}\label{Dscale}
D = \mathop {\arg \max }\limits_r \langle \hat{L}(r)\rangle,
\end{equation}
the location of the maximum value of the time averaged Besag function. The variation of $\langle D \rangle$ will be investigated later when we introduce a modified Vicsek model to account for an external flow.

We also study the morphology of the flock, and we are not aware of any study which has previously done so. To do this one must first compute the shape of the flock. Again there is no unique way to do this and it is clear that results will be somewhat dependent on the method that is used. 
Here our goal is not to perform a detailed analysis of flock morphology, for example to compare with that observed in bird flocks, insect swarms etc. But instead to quantify the change in morphology when an external flow acts on the particles. Hence we choose a computationally efficient method, and apply this consistently in all simulations. We use the particle positions to define an $\alpha$-shape \cite{Edelsbrunner_1983} (a generalisation of the convex hull). Our procedure is the following: Given $N$ particles in a domain of area $L^2$, the expected separation between random points is $\ell=L/\sqrt{N}$. Hence we define a subset of points $\by_i=\{\bx_i : |\bx_i-\bx_{i'}|<\gamma\ell \}$, where $i'$ denotes the closest particle to $\bx_i$.  We take $\gamma=0.5$, which leaves the set of points whose distance to the nearest neighbour is no greater than half the expected separation value. This is to eliminate the contribution of the few points which are not part of the coherent flock (which does occur for the larger values of $\eta$ used in this study).  Note our results are not sensitive to the choice of $\gamma$ (within a sensible range), indeed results are virtually identical if  $\gamma \in [0.25,0.75]$.  The $\alpha$-shape is subsequently constructed from the  points $\by_i$, using a standard package in {\sc MATLAB}. The results of this process can be seen in Fig.~\ref{fig:alpha}. This is then used to compute the morphological properties of the flock using Minkowski functionals, which we discuss next.

Minkowski functionals measure the topological and geometrical features of a given shape or structure. Hadwiger's theorem \cite{klain1997introduction} states that in a $d$-dimensional space, there exist $d + 1$ quantities which completely describe the morphology of a given structure.
In 2D, the three Minkowski functionals of a closed contour are its enclosed area $S$, perimeter $P$, and the Euler characteristic $\chi$ \cite{adler_2007}, but for the purposes of this study it is most convenient to analyse the dimensionless filamentarity
\begin{equation}\label{eq:filament}
F=(P^2-4\pi S)/(P^2+4\pi S)
\end{equation}
By definition, $F \in [0,1]$, with $F =0$ for a circle and $F = 1$ for a (not necessarily straight) line segment. Both $S$ and $P$ can be efficiently computed from the $\alpha$-shape, and so we are able to compute the evolution of $F$ in time, and report an average value for the filamentarity, $\langle F \rangle$, as we did for Besag's functions.
Again we refrain from reporting results here, and present the bulk of results once we have discussed a modification of the model.
Note that the boundary $\partial {\cal S}_\alpha$ of the $\alpha$-shape is a subset of the Delaunay triangulation of ${\cal S}$, hence it is plausible that we over estimate $P$ and hence $F$ due to small triangular artefacts as can be seen in Fig.~\ref{fig:alpha}. Again we emphasise that we are not attempting a detailed analysis of flock morphology, to compare with natural flocks. But to quantify the change in morphology when an external flow acts on the particles using a consistent methodology across all simulations.
\begin{figure}
\begin{center}
\includegraphics[width = 0.4\textwidth]{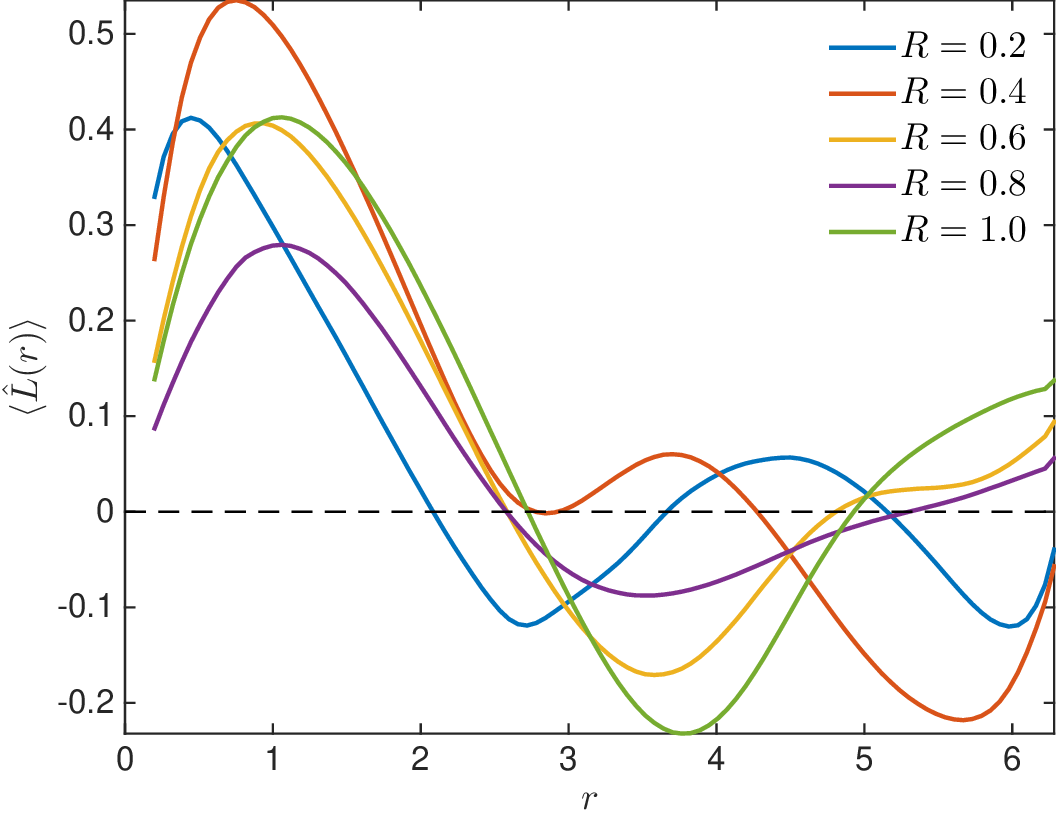}
\caption{(Color online) The time averaged Besag's function $\langle \hat{L}(r) \rangle $, Eq.~(\ref{eq:besag}) plotted as a function of spatial scale $r$ for simulations with $V_f=0.5$. $ \hat{L}(r) >0$ indicates clustering of particles at a given scale, $ \hat{L}(r) <0$ indicates segregation.}
\label{fig:ripley2}
\end{center}
\end{figure}
\begin{figure*}
\begin{center}
\includegraphics[width = 0.4\textwidth]{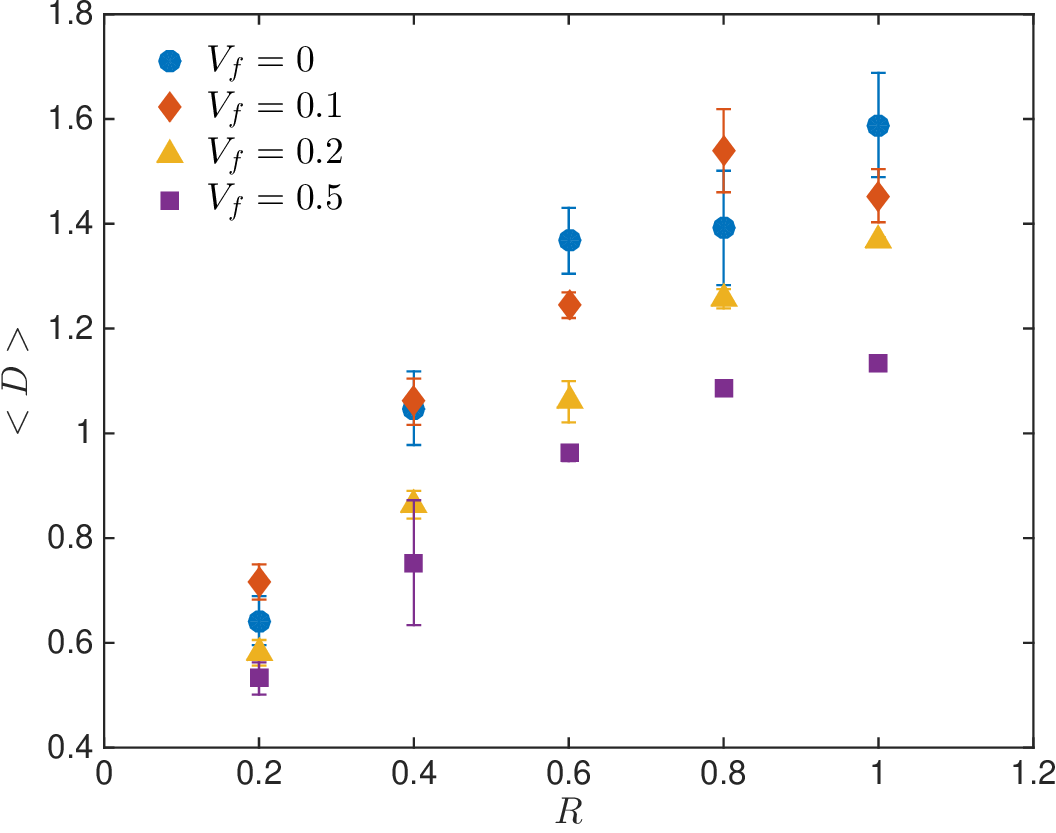}\hspace{1.75cm}
\includegraphics[width = 0.4\textwidth]{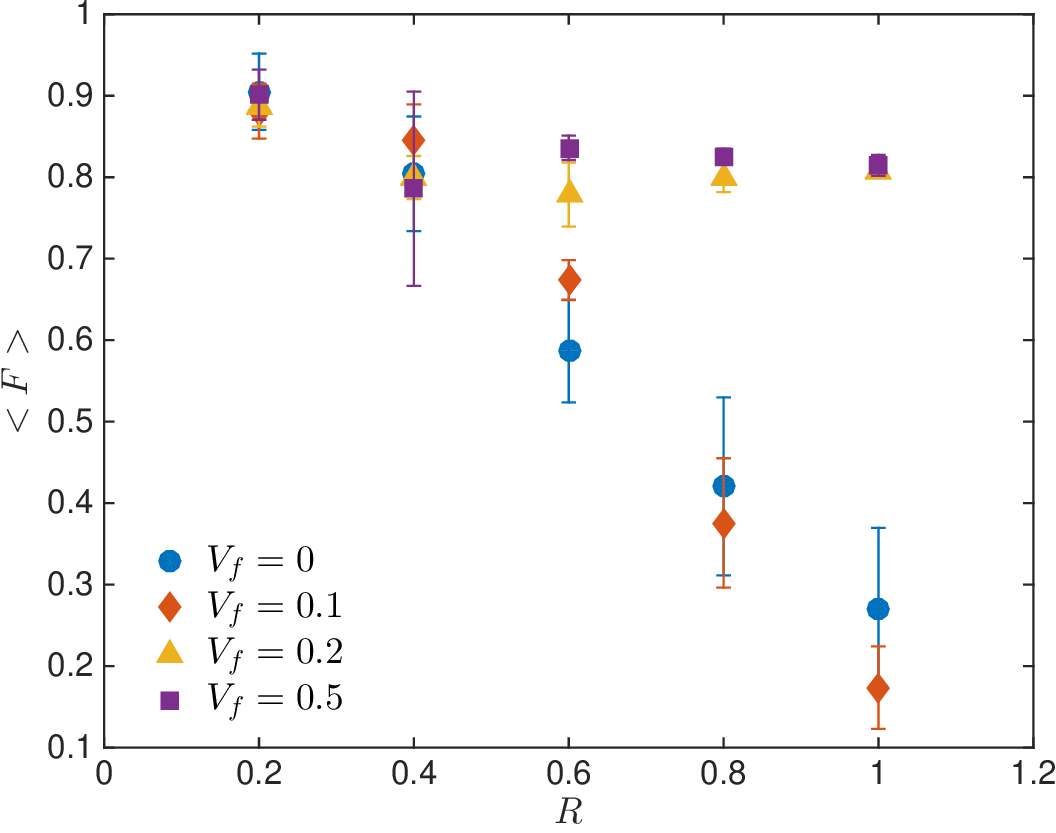}
\caption{(Color online) (left) The time averaged characteristic size of the flock $\langle D \rangle$ (Eq.~(\ref{Dscale})), plotted as a function of the radius of interaction $R$, for varying flow velocities $V_f$. (right) The corresponding plot of the variation of the flock filamentarity $\langle F \rangle$ (Eq.~(\ref{eq:filament})). Note a profound change is observed in the flock morphology ($R\ge 0.6$) when $V_f \ge 0.2$. }
\label{fig:DFvsR}
\end{center}
\end{figure*}

\section{Self-propelled particles in the Taylor Green flow}\label{sec:TGV}
\noindent
We now turn our attention to Vicsek flocks in the presence of external noise, due to fluid motion.
As discussed in section \ref{sec:intro}, we forgo the computational expense, and complexity of a fully turbulent flow. Instead we turn to a steady vortical flow, the Taylor Green (TG) vortex \cite{Taylor1923}, defined as 
\begin{equation}\label{eq:TGV}
\bvf(\bx)=(u_{f},v_{f})=V_f (\sin (x)\cos (y),-\cos (x)\sin (y)),
\end{equation}
where $\bx=(x,y)$. The vorticity field is given by 
\begin{equation}\label{eq:TGVw}
\omega=\nabla \times \bvf=2V_f\sin (x)\sin (y),
\end{equation}
and the flow is incompressible ($\nabla \cdot \bvf=0$).
$V_f$ is a scaling parameter which can be adjusted to modify the relative intrinsic particle speed to that of the background flow. Figure \ref{fig:flow}, displays the vorticity field of the TG flow with the corresponding velocity field.

\begin{figure}
\begin{center}
\includegraphics[width = 0.4\textwidth]{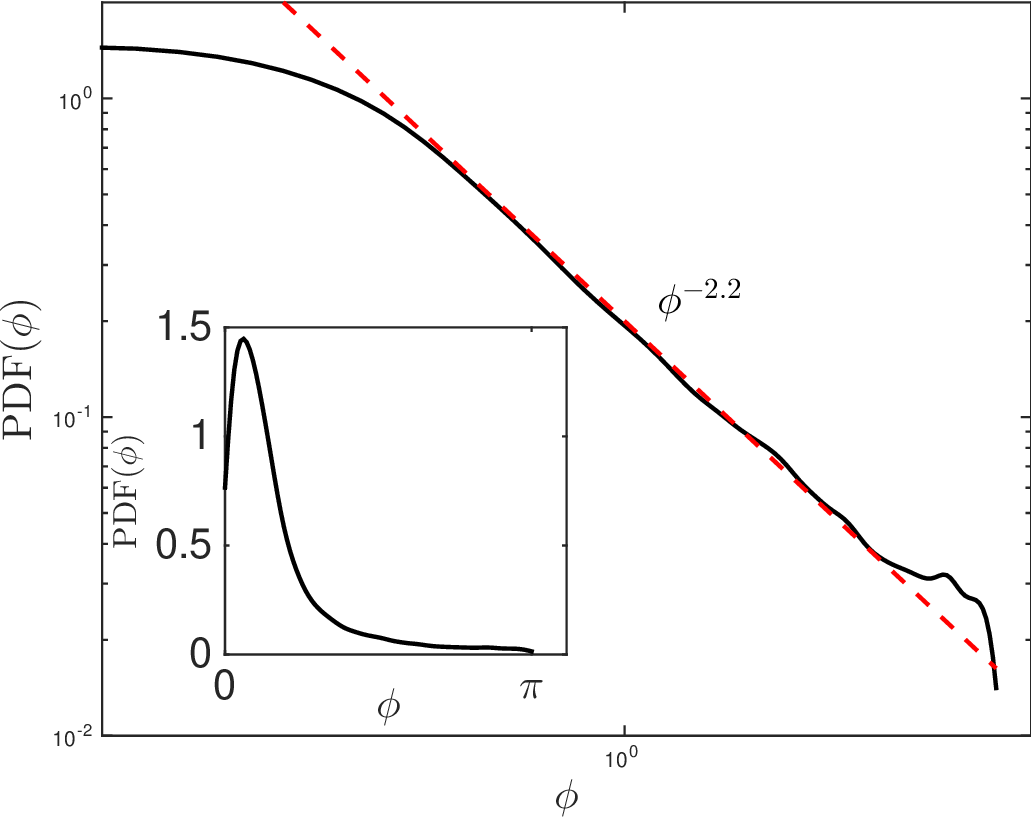}
\caption{(Color online) The probability density function (PDF) of $\phi$, the angle between the flow direction and the particles intrinsic velocity, for the simulation with $V_f=0.2$, $\eta=0.1$, $R=0.6$ . The inset shows the PDF plotted with a linear scale. The main figure shows a logscale plot, with a power law fit ${\rm PDF}(\phi) \sim \phi^{-2.2}$ displayed as a red (dashed) line.}
\label{fig:angle}
\end{center}
\end{figure}

A perhaps overly simplistic approach to modelling the system, would be to simply allow the particles to be advected by the flow, such that the equation of motion for the particles is modified to become

\begin{equation}\label{Eq:velocity2}
\dfrac{{\rm d}\bx_i}{{\rm d} t}=(V\cos(\theta_i)+u_{f}(\bx_i),V\sin(\theta_i)+v_{f}(\bx_i)).
\end{equation}
However, it is not clear to us how Eq.~(\ref{Eq:theta}) can still remain valid. We would assume that `particles' rather than orienting themselves to match the direction the other particles are oriented towards, they would instead orient themselves to the direction of motion of nearby particles. Such an approach was also taken in \cite{Khurana2013}. Hence we arrive at a modified form for the equation used to update $\theta$,
\begin{equation}\label{Eq:theta2}
\theta_i={\rm atan2} \left( \frac{1}{N}\sum_{|\bx_i-\bx_j|< R} \sin \breve{\theta}_j,  \frac{1}{N}\sum_{|\bx_i-\bx_j|  < R} \cos \breve{\theta}_j \right) +\eta \zeta_i,
\end{equation}
where $\breve{\theta}_j={\rm atan2}(V\sin(\theta_j)+v_{f}(\bx_j)),V\cos(\theta_j)+u_{f}(\bx_j))$.
We retain the intrinsic noise, to model the fact that it is unlikely real animals will perfectly align themselves with neighbours within the critical radius. As stated, we believe the justification of noise uncorrelated in space and time is reasonable to model intrinsic noise.
Absent from this modified model is the rotation of particles due to the non-zero vorticity of the flow. The effect of particle rotation would make an interesting starting point to a future study, but we refrain from its inclusion in this work as it would require a substantial modification of the standard Vicsek model.

\begin{figure*}[!ht]
\begin{center}
\includegraphics[width = 0.4\textwidth]{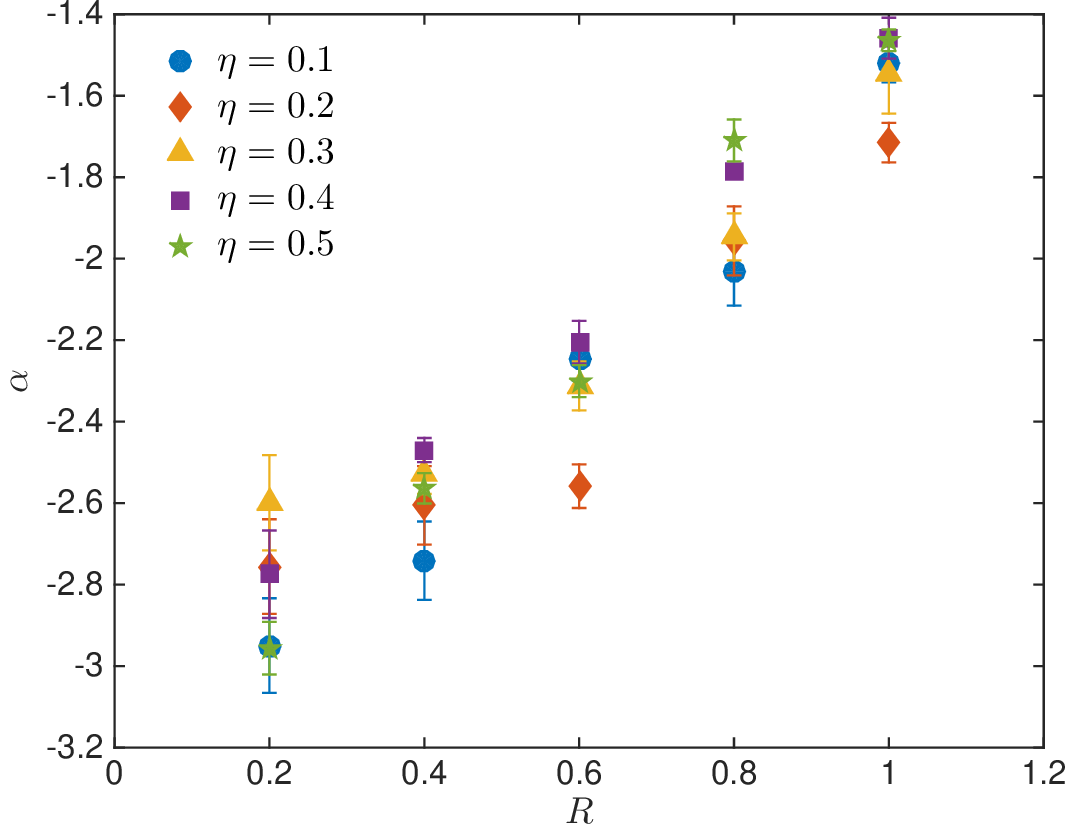}\hspace{1.75cm}
\includegraphics[width = 0.4\textwidth]{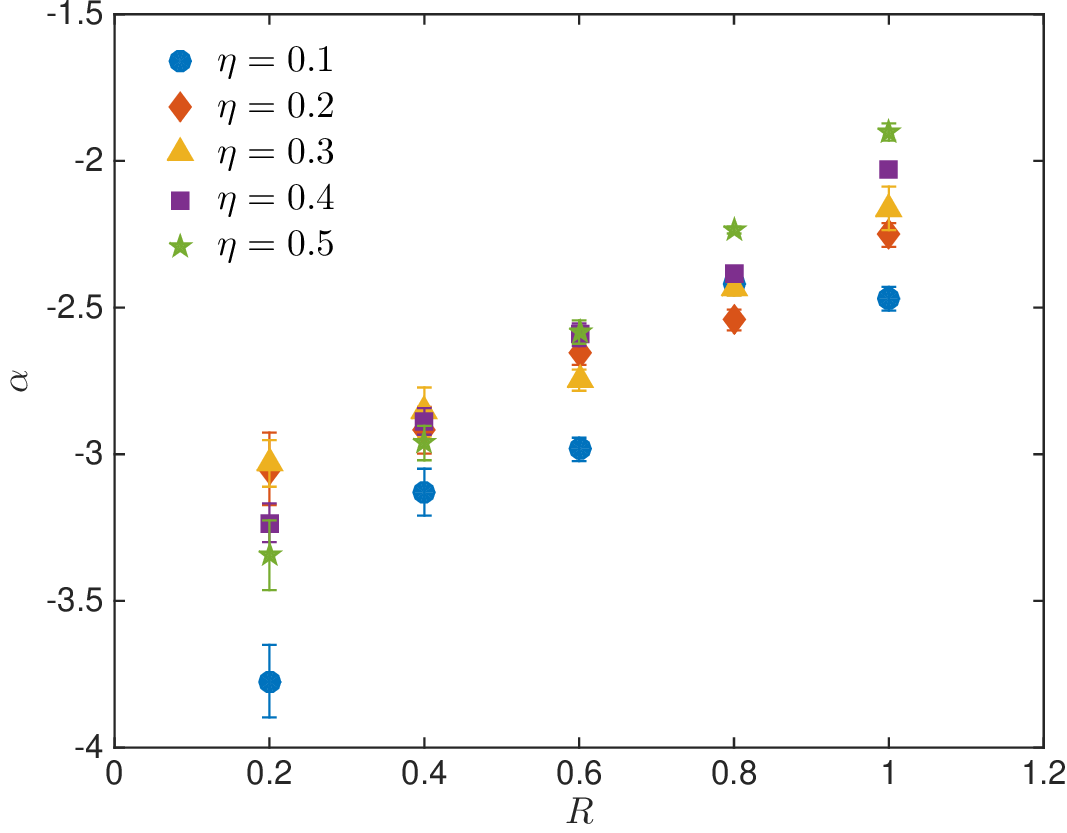}
\caption{(Color online) The power law scaling, $\alpha$, of the probability density function (PDF) of $\phi$ (${\rm PDF}(\phi)=\phi^\alpha$) where $\phi$ is the angle between the flow direction and the particles intrinsic velocity. (left) $V_f=0.2$, (right) $V_f=0.5$, plotted for varying radii of interaction $R$. The slope is clearly dependent on both $R$ and $V_f$, but appears independent of the intrinsic noise $\eta$.}
\label{fig:alphavsR}
\end{center}
\end{figure*}

Particles are evolved according to Eq.~(\ref{Eq:velocity2}), with the particles direction determined using Eq.~(\ref{Eq:theta2}). Domain size and particle number remain unchanged. $\zeta_i$ is drawn randomly at each timestep and we take $\Delta t=0.1$. Unlike for the standard model, we use a $3^{rd}$-order Runge-Kutta time-stepping scheme, to allow for the additional complexity of the flow, however $\theta$ is only updated at the beginning of a timestep. Again simulations ($\eta \in [0,0.5]$, $R \in [0,1]$) are evolved for 200 time units, and we monitor $\psi_{\eta, R}$ to ensure statistics are only computed in a statistically steady value. We consider three values of $V_f=0.1$, $0.2$ and $0.5$, with a fixed value $V=1$ for the particle speed.
Snapshots of the particles for simulations with $\eta=0.1$, $R=0.2, \, 0.8$, $V_f=0.5$ are displayed in Fig.~\ref{fig:vicseck2} along with the vorticity field of the TG flow. It is clear from a visual inspection, particularly for large values of $R$, that the morphology of the flock is dramatically affected by the influence of the flow. We note also that particles tend to be found in areas of low vorticity, this is not seen (and would not be expected) if we consider tracer particles $V=0$. We shall discuss this feature later in the text.

In all simulations we find the statistical properties of the flocks are nearly independent of magnitude of the intrinsic noise, and so for all quantities reported below we perform additional averaging over $\eta$.
Figure \ref{fig:ripley2} shows the averaged Besag's functions (Eq.~(\ref{eq:besag})) for five different values of interaction radius $R$, with $V_f=0.5$. Again a noticeable (although less pronounced when compared with Fig.~\ref{fig:ripley1}) migration of the peak of $\langle \hat{L} \rangle$ to the right is observed. However due to the periodic boundary conditions a secondary peak where $\langle \hat{L} (r)\rangle>0$ for $r>\pi$ can be seen for all values of $R$. This indicates that the flow is forcing the flocks to form an elongated shape, tightly clustered in one direction, but coherent over the scale of the box, as we see in Fig.~\ref{fig:vicseck2}. 

This is also visible if we inspect plots of $\langle D \rangle$ (Eq.~(\ref{Dscale})) against $R$ for the values of $V_f$ considered, Fig.~\ref{fig:DFvsR} (left). A clear reduction in the typical distance of clustering $\langle D \rangle$ is seen (at least for larger values of $R$) as the flow velocity increases. The effect of the flow on the flocks morphology, namely $\langle F \rangle$ (Eq.~(\ref{eq:filament})), can be seen in Fig.~\ref{fig:DFvsR} (right). At low flow speeds ($V_f=0,\, 0.1$) increasing the radius of interaction leads to a dramatic decrease in the filamentarity, as the flock tends to become close to circular. However above some critical value of $V_f$, there is a dramatic transition, and even for large values of $R$ the structures exhibit large filamentarity. It is important to note that this is visible even when the flow speed is only one-fifth of the particles intrinsic swimming speed.

Finally we turn our attention to the relationship between the particles intrinsic direction of motion, and the direction of the local flow. We define 
\begin{equation}
\phi_i=\cos^{-1}\left (\dfrac{ \bv_i \cdot \bvf (\bx_i)}{V | \bvf(\bx_i) |} \right ),
\end{equation}
from this we use kernel density estimation \cite{silverman1986density} to estimate the probability density function (PDF) of $\phi$. This is displayed in Fig.~\ref{fig:angle}. We observe empirically that there is a power law relationship between the angle between the flow and the particles intrinsic velocity, i.e. ${\rm PDF}(\phi) \sim \phi^{\alpha}$. Of course it is then natural to ask if this power  law is universal or depends on $\eta$, $R$ and/or $V_f$. We address this question directly in Fig.~\ref{fig:alphavsR}, for the two larger flow velocities $V_f=0.2, \, 0.5$. It is clear that $\phi$ is independent of the intrinsic noise $\eta$, but depends on both the flow speed $V_f$ and the radius of interaction $R$. The behaviour is as one may naively expect with a much shallower power law relationship for large values of  $R$, and smaller values of $V_f$, where one may expect more scatter between the flow direction and the particles intrinsic direction.

\section{Discussion}\label{sec:discuss}
\begin{figure}[]
\begin{center}
\includegraphics[width = 0.4\textwidth]{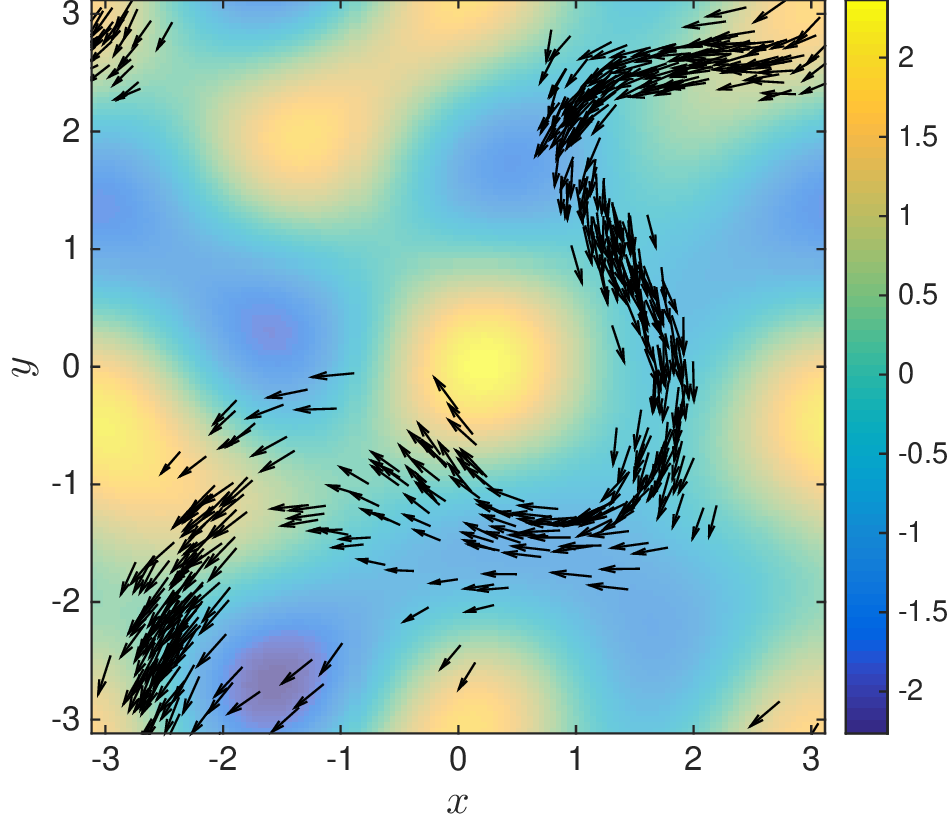}
\caption{(Color online) Snapshots of the Vicsek flock with external fluid forcing given by Eq.~(\ref{eq:KS}),  with $\eta=0.1$, $R=0.8$ and $V_f=0.5$. A pseudocolour plot (made slightly translucent to improve visualisation of the particles) displays the flow's vorticity $\omega$. Arrows indicate the particles direction of motion.}
\label{fig:KS}
\end{center}
\end{figure}
\noindent
The main goal of this work was to understand the effect of an external forcing flow, made up of steady counterrotating vortices, on the standard Vicsek model. To do this we made a slight modification to the model and then compared statistically properties of the flocks, paying particular attention to morphological properties. One of the striking features of the flocks with an external flow $V_f\ge0.2$ is that there is a marked increase in the filamentarity of the flock, when compared to the zero flow case. As is clear in Fig.~\ref{fig:vicseck2} this is due to the fact that particles appear confined to areas of low vorticity. The phenomenon of spatial clustering of independent inertial particles, i.e. subject to Stokes drag, in vortical and turbulent flows is a well explored field \cite{MaxeyWang}. Because the particles posses inertia their flow is compressible, even though the background flow is typically incompressible. Hence large fluctuations of particle concentration are found, and particles tend to collect in elongated sheets (in 3D) on the edges of vortical structures. We see exactly the same phenomena here, not due to a Stokes drag, but because the intrinsic motion of the particles means they do not respond instantaneously to the flow. Hence they are `flung' out of vortical regions and are concentrated in thin ribbons outside of the vortical cells. This is seen to have a profound effect on the morphology of the flock.

In order to investigate the universality of the results presented here we perform one further numerical simulation, with a random, unstructured flow. This is generated from the summation of 3 random Fourier modes in the spirit of the KS model \cite{Fung1992} used in \cite{Khurana2013}. We take a velocity field defined as

\begin{equation}
\bv_f(\bx)=V_f \nu \sum_{m=1}^{3}({\bf A}_m \times \bk_m \cos{\phi_m}
+{\bf B}_m \times \bk_m \sin{\phi_m}),
\label{eq:KS}
\end{equation}

\noindent
where $\phi_m=\bk_m \cdot \bx$, $\bk_m$ are wave vectors, $\mathbf{A}_m$ and $\mathbf{B}_m$ are random unit vectors (orthogonal to $\mathbf{k}_m$) and $\nu$ is a scaling parameter chosen such that the root mean square (RMS) value of $v_f$ is $V_f$. Taking $k_m=m$ creates the random large scale (solenoidal) flow whose vorticity field is displayed in Fig.~\ref{fig:KS}. A simulation of the Vicsek flock in the presence of this random fluid flow is performed taking $\eta=0.1$, $R=0.8$ and $V_f=0.5$; a snapshot of the system in the steady-state is also displayed in Fig.~\ref{fig:KS}.  The elongated structure of the flock visible in the simulations using the TG flow is also found in this random flow, which gives some indication of the robustness of the results presented here.

Of course it is an open question as to the extent to which these results apply to real animal flocks and swarms. However it is both interesting and important to understand how structures in the flow can affect both the shape and fluctuations (which we have not considered here) of the shape of animal flocks. As mentioned earlier in the text we have neglected how rotation of the particles affects their dynamics, which could provide an interesting extension to the model. Natural next steps are to understand the role that temporal fluctuations of the flow play, before turning attention to more realistic flows, such as those generated by direct numerical simulation of the Navier Stokes equation.


\begin{thebibliography}{10}
\bibitem{atan2}
Note the use of the atan2 function, the four-quadrant inverse tangent
  $\tan^{-1}(y,x)$. Two arguments are required, so that the signs of the inputs
  is not lost, essential to return the correct quadrant of the computed angle.
\end{thebibliography}
\end{document}